\documentclass[aps,showpacs,pre,superscriptaddress,twocolumn]{revtex4}
\usepackage{amsfonts}
\usepackage{amsmath}
\usepackage{mathrsfs}
\usepackage{amssymb}
\usepackage{wasysym}
\usepackage{subfigure}
\usepackage{float}
\usepackage{graphicx}
\usepackage{dsfont}

\begin{document}

\title{Topological textures and their bifurcation processes in 2D ferromagnetic thin films}
\author{Jinl\"u Cao}
\affiliation{Department of Physics, Shanghai University, Shanghai 200444, P.R. China}
\affiliation{Qian Weichang College, Shanghai University, Shanghai 200444, P.R. China}
\author{Guo-Hong Yang}
\affiliation{Department of Physics, Shanghai University, Shanghai 200444, P.R. China}
\affiliation{Key Lab for Astrophysics, Shanghai 200234, P.R. China}
\author{Ying Jiang}
\thanks{Corresponding author}
\email{yjiang@shu.edu.cn}
\affiliation{Department of Physics, Shanghai University, Shanghai 200444, P.R. China}
\affiliation{Qian Weichang College, Shanghai University, Shanghai 200444, P.R. China}
\affiliation{Key Lab for Astrophysics, Shanghai 200234, P.R. China}

\begin{abstract}
   In this paper, by the use of the topological current theory, the topological structures and the dynamic processes in thin-film ferromagnetic systems are investigated directly from viewpoint of topology. It is found that the topological charge of a thin-film ferromagnetic system can be changed by annihilation or creation processes of opposite polarized vortex-antivortex pairs taking place at space-time singularities of the normalized magnetization vector field of the system, the variation of the topological charge is integer and can further be expressed in terms of the Hopf indices and Brouwer degrees of the magnetization vector field around the singularities. Moreover, the change of the topological charge of the system is crucial to vortex core reversal processes in ferromagnetic thin films. With the help of the topological current theory and implicit function theorem, the processes of vortex merging, splitting as well as vortex core reversal are discussed in detail.
\end{abstract}

\pacs{75.70.Kw, 75.60.Jk, 03.75.Lm}

\maketitle

\section{Introduction}

Many efforts have been devoted to the magnetic skyrmions and vortices in magnetic thin films due to their potential application in spintronics \cite{2014Topological_Changes, 2013Skyrmion_ground_state, 2004Senthil, 2008Switching_Phenomena, iwasaki-2014, liu-2015, verga-2014, kumar-2014, yamada-2008, noske-2015, kim-2015}. In ferromagnetic systems, skyrmions and vortices are two most common types of topological solitons\cite{1990Kosevich, Book_Topological_Solitons}.

Actually, skyrmions or vortices can naturally arise in 2D magnetic thin films that exceed critical size \cite{1990Cowburn}. Each of these topological textures consists of a core and its surrounding magnetization vector field \cite{Book_Magnetic_Domains, Book_Topological_Solitons, 2014Topological_Changes}. The core is a small region in the 2D medium that has $\vec{M}$ vectors pointing perpendicular to the plane of the film\cite{2000Shinjo}. The magnetic skyrmions are magnetic configurations whose magnetization vectors at spatial infinity are out of plane and antiparallel to the magnetization vectors at the cores, while the magnetic vortices, on the other hand, have in-plane magnetic vectors at spatial infinity.

The sign of the $M_z$ components in the core region gives rise to the polarity $\lambda$ of a topological texture. Apart from the core, the $\vec{M}$ vectors around the core vary from point to point. The total number of times the magnetization vectors wind around the core is the so-called 2D winding number $S$ or vorticity of the topological texture.

Due to their potential applications for information storage and information manipulation\cite{2007Ultrafast_Toggle_Switching_of_Vortex_Cores, 2011magnetic_vortex_memory, 2002_device_applications}, the study of the static and dynamic properties of vortices has recently become one of the main topics in this field \cite{wachowiak, shinjo, choe, park, waeyenberge, guslienko, novosad}. In fact, magnetic vortex cores have indeed been considered as possible candidates for magnetic data storage \cite{kikuchi}. In order to store and manipulate information by utilizing vortex cores, mechanisms for a controlled switching of its orientation are required. Experimental and numerical investigation shows that the magnetic vortex core reversal can be achieved by applying short bursts of alternating out-of-plane magnetic field\cite{waeyenberge, 2007Ultrafast_Toggle_Switching_of_Vortex_Cores} or in-plane current \cite{2014Topological_Changes, 2014_in_plane_current_Skrymion_transition, 2007YLiu}. The reversal of the vortex core is achieved by the creation of a vortex-antivortex pair of whose polarity is opposite to the initial-existing vortex core \cite{2007YLiu} , the newly formed antivortex and the oppositely polarized initial vortex subsequently annihilate, only the newly formed vortex survived, leaving the impression that the polarity of the vortex core is reversed.

Stimulated by the above mentioned progress, in this paper, by utilizing the topological current theory \cite{1999YJ,2000YJ,2004YJ,2009YJ}, we are going to investigate the topological structures and the dynamic processes in a ferromagnetic thin film theoretically from a more fundamental viewpoint, i.e. along the avenue of topology.

The rest of the paper is organized as follows. In section II, by making use of the topological current theory, the inner structure as well as the change of the topological charge of the system is discussed, we find that an annihilation or creation process of a vortex-antivortex pair with opposite polarity leads to an integer leap of the topological charge. In order to investigate the detail information of the vorticity of magnetic vortices and the dynamic processes, such as vortex annihilation, merging, as well as splitting processes, the corresponding topological current is constructed in Section III, and the branch processes as well as the bifurcation properties of the system is discussed. Section IV is devoted to the discussion of the vortex core switching process by the use of the newly constructed topological current.

\section{The topological charge in a 2D thin film and its change }

As is well known, in a ferromagnetic media whose temperature is sufficiently below the Curie point \cite{1996NP}, the magnetization vectors $\vec{M}$ is governed by the Landau-Lifshitz equation \cite{1996NP, 2014Topological_Changes, 2007Rotating_Vortex_Dipoles, 2007YLiu}
\begin{equation}\label{LLE}
\frac{\partial \vec{m}}{\partial t}+ {\vec{m} \times \vec{f}} =\kappa \, \vec{m} \times \frac{\partial \vec{m}}{\partial t}
\end{equation}
in terms of the normalized magnetization vector $\vec{m}=\vec{M}/||M||$, $\vec{f}$ is the effective magnetic field, while the $\kappa$ term serves as the (Landau-Gilbert) dissipation term \cite{1991NP}.

The topological charge of the magnetic configuration of the system is determined by the so-called skyrmion number \cite{1991NP}
\begin{equation} \label{Pontryagin_Index}
Q= \frac{1}{4\pi}\int   \vec{m} \cdot \left(\partial_x  \vec{m} \times \partial _y\vec{m} \right) \,d^2r,
\end{equation}
It is a topological invariant providing $\vec{m}(t,x,y)$ is always well defined in the whole system. Normally, $\vec{m}(t,x,y)$ can be factorized by the so-called stereographic projection to spherical variables $\Theta = \Theta(t,x,y)$ and $\Phi = \Phi(t,x,y)$  as
\begin{equation}\label{Eq_spherical variable}
\left\{
\begin{array}{r@{\;=\;}l}
m_x(t,x,y) & \sin \Theta \cos \Phi,\\
m_y(t,x,y) & \sin \Theta \sin \Phi,\\
m_z(t,x,y) & \cos \Theta,
\end{array}
\right.
\end{equation}
this leads to
\begin{eqnarray}
Q&=& -\frac{1}{4\pi}{\iint} dm_z  d\Phi
\label{QSmz}
\end{eqnarray}
By noting that $S=\frac{1}{2 \pi}\int d\Phi$
is just the vorticity (winding number) of the topological texture, it is not difficult to get that
 \begin{equation}\label{QSmz2}
   Q =  -\frac{1}{2}S\cdot [m_z| _{\rho = +\infty} - m_z| _{\rho = 0}]
 \end{equation}
we then have that the topological charge is expressed as
\begin{equation}\label{QSLambda}
Q=
\begin{cases}
\frac{1}{2} \lambda S, & {\text{ for vortices,}}\\
 \;\; \lambda S,           & {\text{ for skyrmions,} }
\end{cases}
\end{equation}
where $\lambda=m_z|_{\rm core}$ is the polarity of the topological texture.

However, the above parametrization totally neglect the singularities of the system, and cannot be used to investigate the dynamic processes with topological charge changed. Actually, the singularities may exist in 2d lattice magnetic systems \cite{haldane,2004Senthil,2009YJ}. As is pointed out \cite{haldane}, for a magnetic system with lattice structure, when taking continuum limit, singularities of the configuration $\vec{m}$ away from the lattice sites may be allowed, and this will violate the topological conservation in the system \cite{haldane,2009YJ}. Therefore, in order to investigate the the topological structure and the dynamic process of the magnetic configuration generally, the condition for a well-defined $\vec{m}(t,x,y)$ has to be relaxed.

As is known, the unit vector $\vec{m}$ is expressed in terms of the magnetization field as
\begin{equation}\label{phi-def}
m^a=\frac{M^a}{\parallel M\parallel},
\end{equation}
the singularities of $\vec{m}$ directly correspond to the zero points of $\vec{M}(t,x,y)$. The topological charge density in Eq. \eqref{Pontryagin_Index} can now be rewritten in terms of $\vec{M}$ as
\begin{eqnarray}\label{tcDensity4}
q&=&\frac{1}{8\pi} \epsilon^{0\mu\nu}\epsilon_{abc} m^a\partial_\mu m^b \partial _\nu m^c \nonumber \\
&= &\frac{1}{8\pi} \epsilon^{0\mu\nu}\epsilon_{abc} \frac{M^a}{\parallel M \parallel ^3}\partial_\mu M^b \partial _\nu M^c.
\end{eqnarray}
In this expression, the Levi-Civita symbols are used, and for convenience, the space-time variables are rewritten as $z^{\mu}=(t,x,y)$ where the superscript $\mu$ takes the value of 0,1 and 2. By making use of topological current method \cite{2004YJ,2000YJ,1999YJ} and Laplacian Green's function relation, straightforward calculation shows that\cite{2009YJ}
\begin{equation}\label{qt}
\partial _t q= D\left( \frac{M}{z} \right) \delta(\vec{M})
\end{equation}
where $D\left( \frac{M}{z} \right)$ is the Jacobian of $\vec{M}$ and $z^{\mu}$. Eq. \eqref{qt} confirms that when $\vec{M}$ possess no zero points, i.e. when $\vec{m}$ is well defined in the whole space-time, the topological charge is conserved. However, when $\vec{m}$ possesses singular events, the topological charge $Q$ of the system will be changed.

Suppose there are $l$ singularities and the $i$-th one is located at the space-time coordinates $z_i^{\mu}=(t_i,x_i,y_i)$, it is not difficult to verify that
\begin{equation}\label{delta_decompose}
\delta(\vec{M}) = \sum_{i=1}^l c_i \delta(z^{\mu}- z_i^{\mu}), \;\;\; c_i= \frac{\beta_i}{\mid D(\frac{M}{z}) _{\vec{z_i}}\mid}
\end{equation}
where $\beta_i$ is the Hopf index \cite{hopf-brouwer} of $\vec{M}$-field around the $i$-th zero point at $z_i$.

Substituting Eq.\eqref{delta_decompose} back to Eq.\eqref{qt} yields
\begin{equation}\label{qt2}
\partial _t q= \sum_{i=1}^l \beta_i \eta_i \delta(z^{\mu}- z_i^{\mu})
\end{equation}
where $\eta_i = {\rm sgn} D(\frac{M}{z}) _{z_i} $ is the Brouwer degree \cite{hopf-brouwer} of $\vec{M}$-field, and $\beta_i \eta_i$ is the generalized winding number in 2+1-dimensional space-time wrapping around the $i$-th singular event. Therefore, the total change of topological charge is the sum of the all the generalized winding numbers of the singular events emerging in the given time interval
\begin{equation}\label{DeltaQ}
\Delta Q = \int dt \int d^2x \,\partial _t q = \sum_{i=1}^l \beta_i \eta_i,
\end{equation}
due to the second homotopy \cite{AlgebraicTopology} of sphere $\pi_2(\mathbb{S}^2)=Z$, this change is an integer.

Actually, these singular events correspond to annihilation (or creation) processes of vortex-antivortex pairs with opposite polarities in the system. In order to illustrate this point clearly, let us consider an annihilation process of a vortex-antivortex pair with opposite polarities, the vorticity of the vortex (antivortex) is $1$ $(-1)$.

\begin{figure}[h!]
  \centering
  \includegraphics[width=0.8\linewidth]{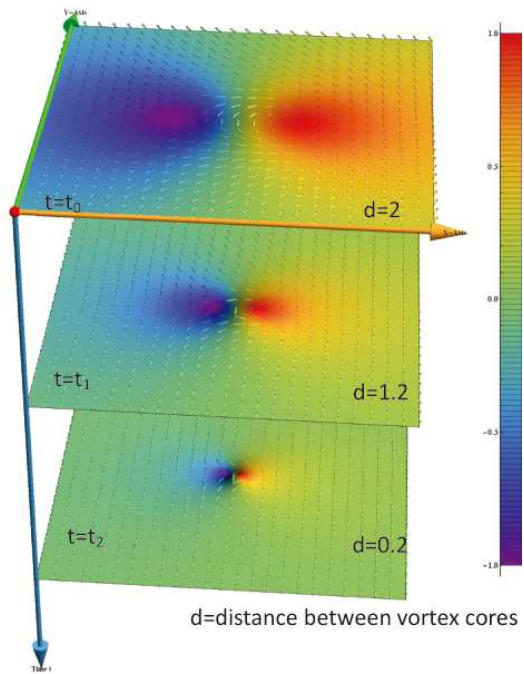}
  \caption{(Color online) The sketch of an annihilation process of a vortex-antivortex pair with opposite polarities. The color gradient bar on the right hand side indicates the value of the $m_z$ component. Each layer of the plane represents a time slice of the magnetic configuration at that moment.}
  \label{opposite_polarities}
\end{figure}

The annihilation process is characterized by the phenomena that cores of the vortex and antivortex collide and merge into one, the merging point is denoted as $z_0=(t_0, x_0, y_0)$. Apparently, when the annihilation process happens, due to the opposite polarity, at the merging point all three component of the magnetization $\vec{M}$ vanishes, leading to a singularity of $\vec{m}$ in the 2D film, according to Eq. \eqref{DeltaQ}, there will be an integer change of $Q$ at $z_0$. This may also be cross-checked by utilizing Eq. \eqref{QSmz2}. From that expression of the topological charge, it is easy to find that the total topological charge before the annihilation is $Q_i=\lambda$ ($\lambda=\pm 1$ is the polarity of the vortex before the annihilation) while the total topological charge after the annihilation is $Q_f=0$, thus $
\Delta Q = Q_{f} - Q_{i} =-\lambda.$ The annihilation process is sketched in Fig. \ref{opposite_polarities}.

Without any difficulty, it can also be seen that a creation process of a vortex-antivortex pair with opposite polarities is also accompanied by $\Delta Q =\pm1$, this was also recognized numerically \cite{2014Topological_Changes}. Moreover, processes with $|\Delta Q|>1$ are also allowed, they correspond to multiple annihilation (creation) events happening simultaneously.

\begin{figure}[h!]
  \centering
  \includegraphics[scale= 0.8]{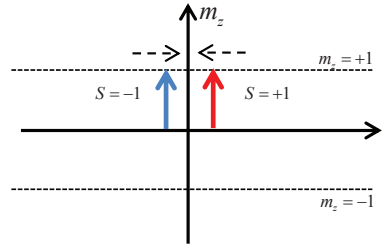}
  \caption{(Color online) The red and blue arrows represent the $m_z$ components of the core of the vortex and the anti-vortex, respectively. Because they have the same polarity, the arrows point in the same direction. As they move closer and closer to annihilation point, $m_z$ component is identical to the previous core.}
  \label{same_polarities}
\end{figure}

For annihilation process of vortex-antivortex pair with the same polarities, however, as shown in Fig. \ref{same_polarities}, the $m_z$ component at the merging point will exactly be the value of the core polarity of each one before the merging, and there is no singular event involved at all and the topological charge $Q$ will keep unchanged in this process.

From above discussion, we see that since $\Delta Q=0$, the annihilation/creation of vortex-antivortex pair parallel polarized cannot be properly reflected in this sense. Even for $\Delta Q\neq 0$ processes, i.e. the annihilation/creation of opposite polarized vortex-antivortex pair, the detailed information also need to be decrypted. Hence, we shall take a step further to construct a topological current which is suitable for discussing the corresponding topological property systematically.

\section{The topological current theory for vortices in 2D thin films}

The above discussion takes the topological structure of the system as a whole, the detail structure of vortices and the corresponding processes can not be exposed by the formula in previous fashion, especially for annihilation (creation) processes of vortex-antivortex pairs with equal polarities. Luckily, it is not necessary to take all three components of the magnetization vectors into account to construct a suitable topological current for vortices in the system. First, it is sufficient to pinpoint the position of the vortex core by solving $m_x=m_y=0$ for $x(t)$ and $y(t)$ at a given time $t$. Second, the primary topological feature $Q$ can be decomposed into two parameters: the vorticity $S$ and the polarity $\lambda$. Among them, the polarity can be treated independently, for it is just the value of $m_z$ at the vortex core which can be directly read out from the configuration. Hence, only the property of the vorticity of the topological texture need to be concentrated on in the following.

\subsection{Constructing 2D topological current}

By choosing $(m_x(t,x,y),m_y(t,x,y))$ as the two components of a vector field  $\vec{\psi}$, i.e.
\begin{equation}\label{phi-mapping}
\left\{
\begin{array}{r@{\;=\;}l}
\psi^1 & m_x(t,x,y)\\
\psi^2 & m_y(t,x,y)
\end{array}
\right. ,
\end{equation}
a topological current can be constructed as follows
\begin{equation}\label{j_mu}
j^\mu = \frac{1}{2\pi} \epsilon ^{\mu\lambda\rho} \epsilon _{ab} \partial _\lambda n^a \partial _\rho n^b,
\end{equation}
where $n^a=\psi^a/\parallel \psi \parallel$. It can be easily verified that $j^{\mu}$ is identically conserved, i.e. \begin{equation}
\label{j_conservation}
\partial _{\mu} j^{\mu}=0.
\end{equation}

By defining the Jacobian determinants $J^{\mu}$ as
\begin{equation}\label{generalized_Jacobian}
\epsilon ^{ab} J ^\mu \left(\frac{\psi}{x} \right) = \epsilon ^{\mu\lambda\rho} \partial _\lambda \psi^a \partial _\rho \psi^b ,
\end{equation}
with the help of the relation between $n^a$ and $\psi^a$ and the 2D Laplacian Green's function relation, a further calculation shows that the topological current in Eq. (\ref{j_mu}) can then be rewritten in terms of $\psi^a$ as
\begin{equation}
j^\mu=   J ^\mu \left(\frac{\psi}{x} \right) \delta(\vec{\psi}) .
\end{equation}
It is obvious that $j^{\mu}$ is nonzero only when $\vec{\psi}=0$. Actually, this expression indicates that all the vortices are located at the zero points of $\vec{\psi}$, i.e. at the zero points of $m^a=(m_x, m_y)$, as we expected. Moreover, the vorticity of each vortex in the system can also be decrypted out of the topological current.

Suppose $\vec{\psi}$ possess $l$ zeroes, and the $i$-th one is located at the spatial-temporal point $\vec{p}_i^*=( x_i^* (t), y_i^* (t))$, by decomposing $\delta(\vec{\psi})$ in terms of $\delta(\vec{x} - \vec{p}_i^*)$, together with the help of the implicit function theorem, when $\vec{p}_i^*$ are regular points, the topological current $j^{\mu}$ can further be rewritten as \cite{1999YJ,2000YJ,2004YJ,2009YJ}
\begin{equation}\label{TTC}
  j^\mu = \sum _{i=1}^l \beta_i \eta_i \delta(\vec{x} - \vec{p}_i^*(t) ) { \left. \frac{J ^\mu \left(\frac{\psi}{x} \right)} { J^0 \left( {\frac{\psi }{x}} \right) }  \right|_{\vec{x} = \vec{p}_i^*} },
\end{equation}
$\beta_i$ is the Hopf index of the vector field $\vec{\psi}$ at zero point $\vec{p}_i^*$ and $\eta_i=\pm 1$ the corresponding Brouwer degree \cite{hopf-brouwer}. Actually, the winding number $W_i=\beta_i\eta_i$ just represents the vorticity $S_i$ of the $i$-th vortex located at $\vec{p}^*_i$, $\eta=1$ for vortex and $\eta=-1$ for antivortex. The density of the vorticity of the topological texture of the system reads
\begin{equation}\label{TTC2}
  \rho = j^0 = \sum _{i=1}^l \beta_i \eta_i \delta(\vec{x} - \vec{p}_i^*(t) ),
\end{equation}
while the velocity of $i$-th vortex(antivortex) moving in the ferromagnetic thin film is
\begin{equation}\label{velocity_of_vortex_core}
  { \left. \frac{J ^\mu \left(\frac{\psi}{x} \right)} { D\left( {\frac{\psi }{x}} \right) }  \right|_{\vec{x} = \vec{p}_i^*} } = v^{\mu},   \mu=1,2
\end{equation}

\subsection{The branch process of topological current}

With the properly constructed topological current $j^{\mu}$ in place,
we can apply the branch process in $\phi$-mapping theory\cite{2000YJ,1999YJ} to provide an understanding of the dynamical behaviors of the vortices at these topologically nontrivial sites with the evolution of time.
For a vortex-antivortex pair located at $\vec{p}_i^*$ and $ \vec{p}_j^*$, i.e. $\vec{p}_i^*=(x_i^*(t), y_i^*(t))$ and $\vec{p}_j^*=(x_j^*(t), y_j^*(t))$ are corresponding solutions of
\begin{equation}\label{phi-mapping3}
\left\{
\begin{array}{r@{\;=\;}l}
\psi^1 (t,x,y) & 0\\
\psi^2 (t,x,y) & 0
\end{array}
\right.
\end{equation}

For the branch process, say, an annihilation event, to occur, the cores of the two vortices have to move closer and closer, until $\vec{p}_i^*$ and $ \vec{p}_j^*$ overlap. As a result, the annihilation event can be characterized by $\vec{p}_i=(x_i(t), y_i(t))$ being a repeated root for Eq.\eqref{phi-mapping3}, thus
\begin{equation}
\left. {J^0\left( {\frac{\psi }{x}} \right)} \right|_{\vec{p}_i^*}
= \left. { {\frac{\partial (\psi^1, \psi^2) }{\partial(x,y)}} } \right|_{\vec{p_i}^*}
=0 .
\end{equation}

Therefore, the spatial-temporal coordinates of the branch process is the solution of
\begin{equation}\label{branch_process}
\left\{
\begin{aligned}
&\psi^1 (t,x,y) = 0 \\
&\psi^2 (t,x,y) = 0 \\
&\psi^3 (t,x,y) \equiv J^0\left( \frac{\psi}{x} \right) = 0
\end{aligned}
\right.
\end{equation}
In fact, there are two possible branch points, namely the limit points and bifurcation points \cite{2000YJ,1999YJ}.
satisfying
\begin{equation}\label{limit_point_condition}
    {\left. J^1(\frac{\psi}{x})\right|}_{(t^*,p_i^*)} \neq 0,
\end{equation}
and
\begin{equation}\label{bifurcation_point_condition}
    {\left. J^1(\frac{\psi}{x})\right|}_{(t^*,p_i^*)} = 0 ,
\end{equation}
respectively. In the following, we are going to discuss the branch processes taking place at these two different type of branch points separately.

\begin{itemize}
  \item Branch Process at Limit Point
\end{itemize}

From implicit function theorem, we can see that when ${\left. J^0(\frac{\psi}{x})\right|}_{(t^*,p_i^*)} = 0$, the solution of the $i$-th singular event $x_i=x_i(t), y_i = y_i(t)$ is not unique. However, in the case of limit point, Eq. \eqref{limit_point_condition} is satisfied, it implies that $x$ and $t$ can change their roles, and the system has a unique solution of the form
\begin{equation}\label{limit_point1}
  t_i=t_i(x),\hspace{1em} y_i = y_i(x)
\end{equation}
in the vicinity of the limit point $(t^*, \vec{p}_i^*)$.
By the use of the third equation in Eq. \eqref{branch_process} and Eq. \eqref{limit_point_condition}, we see from
\begin{equation}\label{velocity_core}
  \frac{dt}{dx} = {\left.\frac{\partial(\psi^1,\psi^2)}{\partial(x,y)} \right/} \frac{\partial(\psi^1,\psi^2)}{\partial(t,y)}
\end{equation}
that the first term in the Taylor expansion for $t_i=t_i(x)$ in the neighborhood of limit point $(t^*, \vec{p}_i^*)$  vanishes, thus, the Taylor expansion up to the second order in the vicinity reads
\begin{equation}\label{limit_point_trajectory}
 t-t_i^* = \frac{1}{2} \left. \frac{d^2t}{(dx^1)^2} \right| _{p_i^*} (x-x_i^{*})^2
\end{equation}
This is a parabola in the $x$-$t$ plane, with each of its branch corresponds to a trajectory of a vortex core hurtling towards the merging point $p_i^*$ at time $t^*$, and an annihilation event occurs.

Generally, $\frac{d^2t}{(dx)^2} |_{p_i^*}$ can be either positive or negative, corresponding to the generation and annihilation of the vortex-antivortex pair, respectively. But in either creation or annihilation case the winding number $S$ (vorticity), due to Eq. \eqref{j_conservation}, should be conserved.

\begin{figure}[h]
  \centering
  \includegraphics[scale= 1]{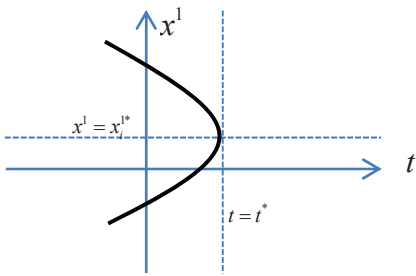}
  \caption{(Color online) A parabola in the $x^1-t$ plane, with each of its branch corresponds to a trajectory of a vortex core moving towards the merging point $p_i^*$ at time $t^*$, and an annihilation event emerges.}
  \label{limit_point_parabola}
\end{figure}

\begin{itemize}
  \item Branch Process at Bifurcation Point
\end{itemize}
Apart from the limit point, what constitutes the larger variety of the vortices' behavior comes from the bifurcation points.
A bifurcation process occurs when the Jacobian ${\left. J^1(\psi/x)\right|}_{(t^*,p_i^*)} = 0$. From Eq. \eqref{velocity_core}, one can easily see that the velocity of the vortex core is undetermined. Therefore, it is easy to picture the situation that the trajectory of the core at the bifurcation point will diverge into multiple curves, because the function relationship between $t$ and $x$ is not unique in the vicinity of the bifurcation point $(t^*, \vec{p}_i^* )$.

Because the Jacobian $J^0(\psi/x)=0$, the rank of the Jacobian will be smaller than 2. As in our case, for practical reasons, we consider the rank to be 1. Assume $J^0_1 = \psi^1_2\equiv \frac{\partial \psi^1}{\partial y}$ is nonzero in the Jacobian matrix $J^0(\frac{\psi}{x})=0$.
Then the implicit function theorem ensures that there exists one and only one function relation solved from $\psi^1(t,x,y) = 0$, that is
\begin{equation}\label{f_in_bifurcation}
  y = f(x,t)
\end{equation}
We denote the partial derivatives as $f_1 = \frac{\partial f}{\partial x}, f_t=\frac{\partial f}{\partial t},
f_{11}=\frac{\partial^2 f}{\partial x^2 }, f_{1t}=\frac{\partial^2 f}{\partial x \partial t}, f_{tt}=\frac{\partial^2 f}{\partial t^2}$.
From Eqs. \eqref{branch_process} and \eqref{f_in_bifurcation}, we have
\begin{equation}\label{phi_in_bifurcation}
  \psi^1= \psi^1 (t, x, f(x,t)) = 0
\end{equation}
which give
\begin{align}
  \label{1st_derivatives_bifurcation1} &\frac{\partial \psi^1}{\partial x} = \psi_1^1 + \frac{\partial \psi^1}{\partial f} \frac{\partial f}{\partial x}=0 \\
  \label{1st_derivatives_bifurcation2} &\frac{\partial \psi^1}{\partial t}   = \psi_t^1 + \frac{\partial \psi^1}{\partial f} \frac{\partial f}{\partial t}=0
\end{align}
By differentiating Eqs. \eqref{1st_derivatives_bifurcation1} and \eqref{1st_derivatives_bifurcation2} with respect to $x$ and $t$, applying the Gaussian elimination method, the second order derivatives $f_{11},f_{1t}$ and $f_{tt}$ can be found. The above results are made out quite independent of the last component $\psi^2(t,x,y) $. In order to find the different trajectories in the bifurcation process, the Taylor expansion of the remaining component $\psi^2(t,x,y)$ in the neighborhood of $(t^*, \vec{p}_i^*)$ should be studied. By substituting Eq. \eqref{f_in_bifurcation} into $\psi^2(t,x,y)$, we define
\begin{equation}\label{F_in_bifurcation}
  F(x,t)= \psi^2 (t, x, f(x,t))
\end{equation}
By using the condition for branch process, i.e. Eq. \eqref{branch_process}, we have
\begin{equation}\label{F=0}
F(t^*, \vec{p}_i^*)=0.
\end{equation}
Besides, from Eq. \eqref{F_in_bifurcation}, the 1st order derivatives are
\begin{equation}\label{1st_derivatives_F}
\frac{\partial F}{\partial x^1} = \psi_1^2 + \frac{\partial \psi^2}{\partial f} f_1=0 , \hspace{1 em}
\frac{\partial F}{\partial t}   = \psi_t^2 + \frac{\partial \psi^2}{\partial f} f_t=0
\end{equation}
By utilizing Eqs. \eqref{1st_derivatives_bifurcation1} and  \eqref{1st_derivatives_bifurcation2}, the Jacobian determinant $J^0( \psi /x )$   in the third equation of Eq. \eqref{branch_process} can be rewritten as
\begin{equation}
\frac{\partial F}{\partial x} J^0_1 |_{t^*,\vec{p}_i^*}=0.
\end{equation}
As $J^0_1 \neq0$, the above equation gives
\begin{equation}\label{partial_F_1}
{\left. \frac{\partial F}{\partial x} \right|}_{t^*,\vec{p}_i^*} = 0
\end{equation}
Similarly,
\begin{equation}\label{partial_F_t}
{\left. \frac{\partial F}{\partial t} \right|}_{t^*,\vec{p}_i^*} = 0
\end{equation}
The second order partial derivatives of $F$, can be calculated from Eq.\eqref{1st_derivatives_F}, and we further denote them by
$A={\left. \frac{\partial^2 F}{(\partial x )^2} \right|}_{t^*,\vec{p}_i^*} ,
B= {\left. \frac{\partial^2 F}{\partial x \partial t} \right|}_{t^*,\vec{p}_i^*} ,
C= {\left. \frac{\partial^2 F}{\partial t^2} \right|}_{t^*,\vec{p}_i^*} $.

\begin{figure}[h]
  \centering
  \includegraphics[width=0.7\linewidth]{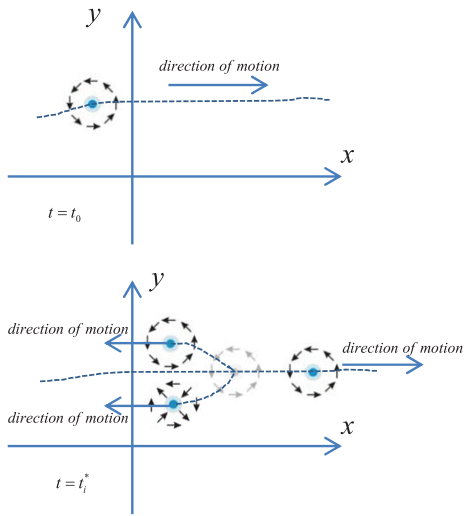}
  \caption{(Color online) A sketch of a single vortex splitting into multiple vortices at the bifurcation point.}
  \label{bifurcation_sketch}
\end{figure}

By virtues of Eqs. \eqref{F=0}, \eqref{partial_F_1}, and \eqref{partial_F_t} the Taylor expansion of $F(x,t)$ in the vicinity of bifurcation point is
\begin{equation}\label{expansion_at_bifurcation}
  A (x - x_i^{*})^2 +2B (x - x_i^{*})(t-t^*) + C(t-t^*)^2 = 0
\end{equation}
Divide Eq. \eqref{expansion_at_bifurcation} by $(t-t^*)$, and take the limit of $t\rightarrow t^*, x \rightarrow x_i^{*}$, we get
\begin{equation}\label{bifurcation_trajectory1}
  A (\frac{dx}{dt})^2 +2B \frac{dx}{dt} + C = 0
\end{equation}
or
\begin{equation}\label{bifurcation_trajectory2}
  C(\frac{dt}{dx})^2 +2B\frac{dt}{dx} +A=0
\end{equation}
while the remaining component $\frac{dy}{dt}$ is determined from $y = f(x,t)$, which yields
\begin{equation}\label{bifurcation_trajectory3}
  \frac{dy}{dt}=f_1 \frac{dx}{dt} +f_t
\end{equation}
The above equation shows that the second components of the trajectory at the bifurcation point can diverge as well, for $\frac{dx}{dt}$ may choose different values. This can be used to explain phenomena which is categorized as a bifurcation process. For instance, a case when two vortices coming into a merging point and then diverge, a single vortex due to certain reasons spilt into several, a number of vertices merge into one at the bifurcation point, etc. Fig. \ref{bifurcation_sketch} depicts one of the possible scenarios described above.

\section{The vortex core switching in a 2D ferromagnetic thin-film}

With the above established topological current in hand, we can now investigate the vortex core switching process in detail theoretically. The vortex core reversal processes have been numerically simulated successfully \cite{2007YLiu, hertel}. Immediately after a field pulse \cite{hertel} or a polarized current pulse \cite{2007YLiu} is applied to a 2D nanodisk possessing a vortex with positive polarity,  i.e. $\lambda = m_z|_{\rm{core}}=+1$, the magnetization of the nanodisk is heavily distorted and a magnetic bubble \cite{2009magnetic_bubbles,1996NP,1991NP} with negative polarity is formed, leading to a creation of a vortex-antivortex pair. The newly formed antivortex and the oppositely polarized initial vortex subsequently annihilate, leaving the new vortex oppositely polarized with respect to the initial one.

Due to its potential application in information storage and manipulation, it is worth to investigate the topological property of the system in the vortex core switching processes.

Although multiple switching process can be discussed in the same fashion, for simplicity and clarity, we only set out to investigate the single switching process. By the use of the topological current theory, we will show that the first step in the core switching process, i.e. the creation of a vortex-antivortex pair, can be classified as a $\Delta Q = 0$ bifurcation process while the latter subprocess is a $\Delta Q=1$ vortex-antivortex pair annihilation process at a limit point.

In order to study the detail topological information of the dynamic process, an initial state need to be constructed to describe the magnetic configuration of the disk right after the magnetic bubble induced by the pulse. Then, by plugging it back into Landau-Liftshitz-Gilbert equation as an initial state, the property of the vortex system will be revealed.

For simplicity but without loss of generality and the topology of the vortex state, the trial state may be constructed by the use of the general static solutions of the Landau-Lifshitz equation, that is the well-known Belavin-Polyakov instanton \cite{belavin-77, 1991NP, 2007Rotating_Vortex_Dipoles}. This is a static solution of the simplified Landau-Lifshitz equation, which only includes the exchange interactions \cite{1991NP}:
\begin{align}
  \frac{\partial \vec{m}}{\partial t}= \vec{m} \times \Delta \vec{m}, \hspace{1.0em} \mid\vec{m} \mid =1, \nonumber
\end{align}
By re-parameterizing the magnetization vector $\vec{m}$ with the complex quantity
$$\Omega = \frac{m_x+im_y}{1+m_z}$$
and using the complex variables $z=x+iy$, the LLE reduces to
\begin{align}
  \frac{i}{4} \frac{\partial \Omega}{\partial t} +\Omega_{\overline{z}z}= \frac{2\overline{\Omega}\Omega_{\overline{z}}\Omega_z}{1+\Omega \overline{\Omega}} .
\end{align}
All solutions of this equation can be constructed from the following four types of basic solutions which, by properly choosing the coordinate system, take forms as
\begin{align}
  &\Omega_1=\left( \frac{z}{a_1}\right), \hspace{1em} \Omega_2=\left(\frac{a_2}{z}\right), \nonumber \\
  &\Omega_3=\left(\frac{\overline{z}}{a_3}\right), \hspace{1em} \Omega_4=\left(\frac{a_4}{\overline{z}}\right),  \nonumber
\end{align}
in this expression, the corresponding core for each type of basic configuration is exactly located at the original point of the coordinate system, and $a_i$ represents the relative size of each type of vortex. It is not difficult to check the polarity $  \lambda = \mathop{ {\rm lim}}_{\rho\rightarrow0} m_z = \mathop{ {\rm lim}}_{\rho\rightarrow0} \frac{1-\Omega\overline{\Omega}}{1+\Omega\overline{\Omega}}$ of each type and find
\begin{align}
  \lambda_{\Omega_1} = 1, \hspace{1em} \lambda_{\Omega_2}=-1, \nonumber \\
  \lambda_{\Omega_3} = 1, \hspace{1em} \lambda_{\Omega_4}=-1. \nonumber
\end{align}
The vorticities for $\Omega_1$ and $\Omega_2$ are $1$, while both $\Omega_3$ and $\Omega_4$ have vorticity of $-1$.

The pulse creates a magnetic bubble with opposite polarity with respect to the initial-existing vortex, i.e. $\lambda=-1$, only $\Omega_2$ and $\Omega_4$ are suitable candidates to construct the corresponding magnetic configuration of the induced magnetic bubble. Moreover, due to the conservation property of the topological current, i.e. $\partial _{\mu} j^{\mu}\equiv 0$, the 2D winding number of the system should be conserved, hence the configuration for the magnetic bubble can only be a superposition of $\Omega_2$ and $\Omega_4$ which reads
\begin{equation}
\Omega=\Omega_2 \Omega_4=\left(\frac{a}{\rho} \right)^2
\end{equation}
where $\rho=z \overline{z}=x^2+y^2$, $a^2=a_2 a_4$ representing the relative size of the magnetic bubble \cite{1991NP}.

\begin{figure}[h]
  \centering
  \includegraphics[width= 0.9\linewidth]{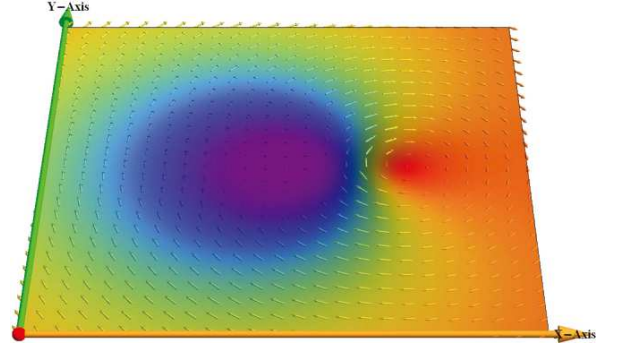}
  \caption{(Color online) The magnetic configuration in Eq. \eqref{Total_configuration_Vortex_DIY}}
  \label{fig_switching_process_stage_ONE}
\end{figure}

By taking the initially existing vortex with $\lambda = 1$ and $S=1$ into account, the total configuration of the entire disk is $\Omega = \Omega_1 \Omega_2 \Omega_4$, for simplicity but without loss of topological information, the magnetic configuration possessing the same topological feature as in the simulation \cite{2007YLiu} reads
\begin{equation} \label{Total_configuration_Vortex_DIY}
  \left\{
\begin{aligned}
  m_x &= -\frac{2 b y \left(\frac{a^2}{x^2+y^2}\right)^n}{\left((d-x)^2+y^2\right) \left(\frac{a^2}{x^2+y^2}\right)^{2 n}+b^2}  \\
  m_y &= \frac{2 b (x-d) \left(\frac{a^2}{x^2+y^2}\right)^n}{\left((d-x)^2+y^2\right) \left(\frac{a^2}{x^2+y^2}\right)^{2 n}+b^2}  \\
  m_z &= \frac{b^2-\left((d-x)^2+y^2\right) \left(\frac{a^2}{x^2+y^2}\right)^{2 n}}{\left((d-x)^2+y^2\right) \left(\frac{a^2}{x^2+y^2}\right)^{2 n}+b^2}
\end{aligned}
  \right.
\end{equation}
where $d$ is the distance between the initial vortex and the newly formed magnetic bubble, $b$ reflects the size of the core of the initial vortex. This configuration is shown in Fig. \ref{fig_switching_process_stage_ONE}.

Together with Eqs. \eqref{phi-mapping} and \eqref{generalized_Jacobian}, for the above magnetic configuration£¬it is found that $J^0(\psi/x)|_{\rho \rightarrow 0}=0$, indicating that the system will undergo a branch process at the instant the pulsed current induces the zero-winding number vortex. To determine the type of the branch process, the dynamical equation£¬which governs the vortex evolution, should be brought into the fold, that is the Landau-Lifshitz-Gilbert equation
 \begin{equation}
   \frac{\partial \vec{m} }{ \partial t} = -\gamma \vec{m} \times \vec{H}_{\textrm{e}} + \alpha \vec{m} \times \frac{\partial \vec{m} }{ \partial t}
   - (\vec{u} \cdot \bigtriangledown) \vec{m} + \beta \vec{m} \times [(\vec{u} \cdot \bigtriangledown)\vec{m}], \nonumber
 \end{equation}
 where $\alpha$ is the Gilbert damping constant, $\beta$ is the dimensionless parameter describing the nonadiabatic process, and
 \begin{equation}
 \vec{H}_{\textrm{e}} = \Delta \vec{m} + \gamma' (\partial_x (\partial_{\mu} m_{\mu}) ,\partial_y (\partial_{\mu} m_{\mu}),0)
 \end{equation}
is the effective field containing the exchange interactions and the magnetostatic energy.
By substituting the detailed expression of the magnetic configuration in Eq. \eqref{Total_configuration_Vortex_DIY} into above equations, direct calculation shows that
 \begin{align}
    {\left.\frac{\partial \vec{m} }{ \partial t} \right|}_{\rho\rightarrow0,t=0}
       &= \frac{4 b \gamma (\gamma ' + 2)}{a^2 d (1+\alpha ^2)} \, (1,-1,0),
 \end{align}
Further calculation leads to
\begin{equation}\label{J1}
  {\left. {J^1 \left( \frac{\psi}{x} \right)} \right|}_{\rho\rightarrow0,t=0} = {\left. \frac{\partial(m_x,m_y)}{\partial(t,y)} \right|}_{\rho\rightarrow0,t=0} = 0
\end{equation}
Now, with both $D\left(\frac{\psi}{x}\right)=0$ and $J^1\left(\frac{\psi}{x}\right)=0$ at the branch point, according to the the discussion in preceding section, we see that immediate after the creation of the magnetic bubble the system will undergo a bifurcation process. Direct application of the bifurcation process in topological current theory yields the differential equations that governs the trajectories of the vortex cores in the neighborhood of the bifurcation point, which is
\begin{equation}\label{bifurcation trajectory_LYWsimulation}
\left\{
\begin{aligned}
&A \left( \frac{dx}{dt} \right)^2 + 2B \frac{dx}{dt} +C = 0 \\
&\frac{dy}{dt} = \frac{\partial f}{\partial x} \frac{dx}{dt} +\frac{\partial f}{\partial t}
\end{aligned}
\right.
\end{equation}
where $y= f(x,t)$, and the two solutions of the differential equations will corresponds to the equation of motion of the two vortex cores split from the zero-winding-number magnetic bubble in the vicinity of the bifurcation point. Moreover, according to the conserved 2D topological current, the newly-generated vortices should have equal but opposite vorticity, i.e. these two vortices should be a vortex-antivortex pair. The formation of vortex-antivortex pairs after application of short current pulses is consistent with recent experimental observations \cite{klaui}.

\begin{figure}[h!]
  \centering
  \includegraphics[width= 0.9\linewidth]{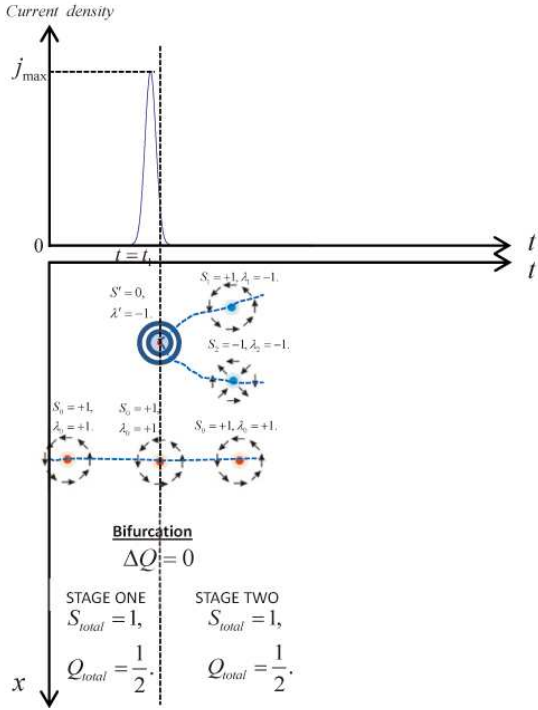}
  \caption{(Color online) A sketch of process of the magnetic bubble with negative polarity created by a pulsed current and then splits into a vortex-antivortex pair. During this process, the total topological charge, i.e. the skyrmion number of the system, keeps unchanged, $\Delta Q=0$ }
  \label{fig_Bifurcation_Switching1}
\end{figure}

After the bifurcation process, the system possesses totally three vortices, one is the initially existing vortex with $\lambda=+1$, the other two are the vortex-antivortex pair with $\lambda =-1$, at this stage the topological current describing the system now reads
\begin{align}\label{TC_1}
  j^\mu =& \sum _{i=0}^{l-1} \beta_i \eta_i \delta(\vec{x} - \vec{p}_i^*(t) ) { \left. \frac{J ^\mu \left(\frac{\psi}{x} \right)} { D\left( {\frac{\psi }{x}} \right) }  \right|_{\vec{x} = \vec{p}_i^*} } \nonumber \\
   =
    & \delta(\vec{x} - \vec{p}_0^*(t) ) { \left. \frac{J ^\mu (\psi/x )} { D(\psi/x ) }  \right|_{\vec{x} = \vec{p}_0^*} } \nonumber \\
   +& \delta(\vec{x} - \vec{p}_1^*(t) ) { \left. \frac{J ^\mu (\psi/x )} { D(\psi/x ) }  \right|_{\vec{x} = \vec{p}_1 ^*} } \nonumber \\
   -& \delta(\vec{x} - \vec{p}_2^*(t) ) { \left. \frac{J ^\mu (\psi/x )} { D(\psi/x ) }  \right|_{\vec{x} = \vec{p}_2 ^*} }
\end{align}
The core positions of the vortices are located at $\vec{p}_i^*(t)$, respectively, and the trajectories of the vortices in the 2D nanodisk can be obtained. If we denote the time when the first branch process event happens, i.e. the bifurcation discussed above, to be $t_1$, then one can arrive at the sketch in Fig. \ref{fig_Bifurcation_Switching1} depicting the motion of $x$ component of the vortex core as a function of time.

In the above bifurcation process, the trajectories of the newly generated vortex and the antivortex diverge and the splitting vortices drift apart. In order to investigate the following dynamic process, the interactions between the newly generated vortex (antivortex) and initially existing vortex then need to be taken into account.

As is known \cite{2004Lara}, the effective interaction between vortices in the magnetic environment is
\begin{equation}
\vec{F}_{ij} = \frac{2\pi J S_iS_j}{X_{ij}}\hat{X}_{ij}
\end{equation}
where $J$ is the magnetic coupling constant in the magnetic system (for our ferromagnetic thin-film system, $J>0$), $\hat{X}_{ij}$ is the unit vector pointing from the center of $i$-th vortex to the center of $j$-th vortex, and $X_{ij}$ the distance between these two vortices, $S_i$ and $S_j$ are the corresponding vorticities of the two vortices. It is then easily verified that the interaction between the newly generated antivortex and the initially existing vortex is attractive while the interaction between the newly generated vortex and the initially existing vortex is repulsive.

\begin{figure}[h!]
  \centering
  \includegraphics[width= 0.9\linewidth]{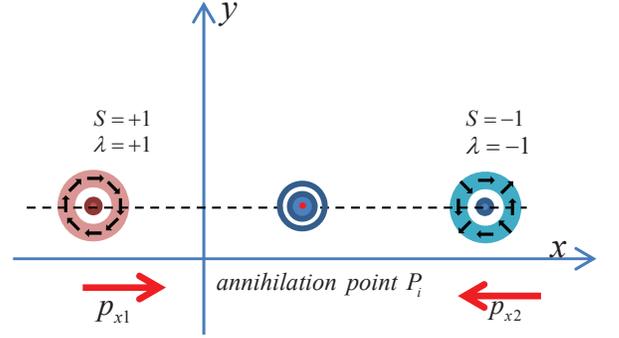}
  \caption{(Color online) A sketch of a head on collision of a vortex-antivortex pair with $\Delta Q = -1$ }
  \label{fig_opposite_polarity_limit_point}
\end{figure}

The attractive interaction between the antivortex and the vortex may lead to a collision and merge. During the process, due to the conservation of the vortex momentum \cite{1991NP}, in the center-of-mass reference frame, the total momentum $\vec{p}=\vec{p}_{\rm vortex}+\vec{p}_{\rm antivortex}$ should be zero all the time during the process. Meanwhile, for the vortex-antivortex pair, the total vorticity is zero, the corresponding virial relation reads \cite{1991NP} $\vec{v}\cdot \vec{p}=-E$, where $E$ is the total energy of this vortex antivortex pair. By properly choosing the reference coordinate as shown in Fig. \ref{fig_opposite_polarity_limit_point}, also taking into account that the total momentum is zero while the total energy is finite, we then have $1/v_x=0$. Together with Eq. \eqref{velocity_of_vortex_core}, we finally draw the conclusion that at the merging point $P^*$ we have $J^0(\psi/x)|_{P^*}=0$ while $J^1(\psi/x)|_{P^*} \neq 0$, indicating that the merging point in this stage is a limit point. By the use of the topological current theory at limit point which is discussed in preceding section, the relation between $x$-components of the vortex-core positions and $t$ is
\begin{equation}
t-t^* = \frac{1}{2} \left. \frac{d^2t}{dx^2} \right| _{P^*} (x-x^*)^2
\end{equation}
which is a parabola in the $x$-$t$ plane.

From Fig. \ref{limit_point_parabola}, it can be recognized that the upper and lower branch of the parabola correspond to the $x$-components of trajectories of the vortex and antivortex cores respectively. More over, if $\left. \frac{d^2t}{dx^2} \right| _{P^*} <0$,  then the opening of the curve is to the left, therefore with the passing of time the position of the vortex and antivortex core will become closer and closer to the limit point, and both vortices cease to exist the moment after $t^*$, indicating that an annihilation event takes place at $P^*$. On the other hand, if $\left. \frac{d^2t}{dx^2} \right| _{P^*} >0$, then this limit process will correspond to the creation of vortex-antivortex pair. Actually, it is not difficult to verify that
\begin{equation}
{\rm sgn} (\left. \frac{d^2t}{dx^2} \right| _{P^*}) = {\rm sgn} (\frac 1{v_x(x-x^*)}| _{P^*}).
\end{equation}
In our case, due to the attractive force between the two vortices of the vortex-antivortex pair, the sign of $v_x$ is always opposite to the sign of $(x-x^*)$, therefore, $\left. \frac{d^2t}{dx^2} \right| _{P^*} <0$, indicating that the branch process at this limit point is a vortex-antivortex annihilation process.

\begin{figure}[h]
  \centering
  \includegraphics[width= 0.9\linewidth]{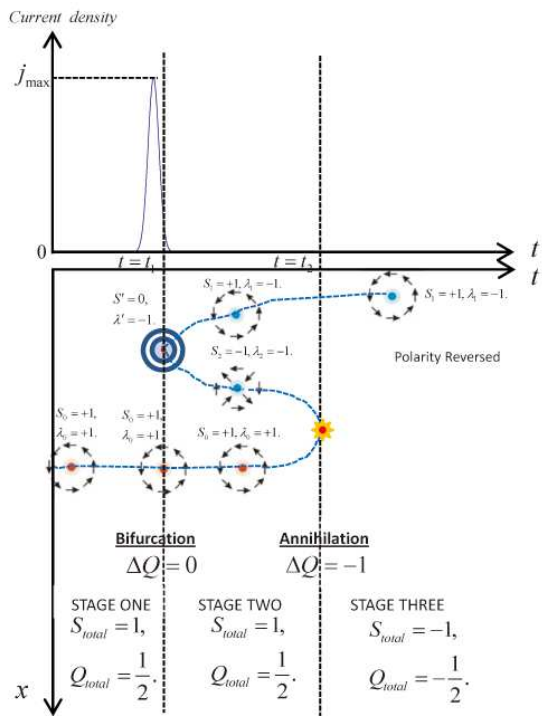}
  \caption{(Color online) A sketch of the system's development throught the switching process.
   }
  \label{fig_Bifurcation_Switching2}
\end{figure}

With the final piece in place, a complete sketch of the single switching process of magnetic vortex core in a ferromagnetic thin film is drawn in Fig. \ref{fig_Bifurcation_Switching2}. Long before a pulse is applied to the magnetic thin film, there is only one vortex with vorticity $S_0=1$ and polarity $\lambda=1$ in the system. At time $t_1$, a pulse is applied which induces a negative polarized magnetic bubble with zero winding number. Immediately after the creation of such a magnetic bubble, it quickly bifurcates into a vortex-antivortex pair oppositely polarized with respect to the initially existing vortex. Since so far $\vec{m}$ is well defined in the whole system, the topological charge of the system keeps unchanged during the first sub-process of the vortex-core switching. Due to the attractive interaction between the newly formed antivortex and the initial vortex, they annihilate each other at time $t_2$, leaving the newly formed negatively polarized vortex survived in the system, and the vortex-core switching is accomplished. Clearly, the topological charge of the system is $Q=1/2$ before the annihilation event while $Q=-1/2$ after that, hence during the second sub-process of vortex-core switching, $\Delta Q=-1$, indicating that there is a space-time singular event involved in the process.

\section{Conclusion}

In summary, by the use of the topological current theory, we have discussed the topological structure and dynamic processes in a thin-film ferromagnetic system.

We found that the topological charge of the magnetic system can be changed when space-time singularities of normalized magnetization field of the system are allowed, and the variation of the topological charge can only be integer which is exactly the total wrapping number of the corresponding singular events. We found that such integer leap of the topological charge of the system is always associated with annihilation or creation processes of vortex-antivortex pair with opposite polarities, these processes can be looked upon as space-time skyrmions or monopole events in the system.

In order to investigate the topological properties of magnetic vortices in 2D ferromagnetic thin films, we have constructed a topological current out of the in-plane components of the normalized magnetization vector field of the system. We found that the vorticity of a vortex can further be expressed in terms of the Brouwer degree and Hopf index of the normalized magnetization vector field around the vortex core. We found that the creation (annihilation) of vortex-antivortex pairs correspond to the topological current branch processes at limit points, while the merging and splitting of vortices can be explained by the topological current branch theory at bifurcation points. By the use of this topological current theory, the vortex-core switching process in a 2D ferromagnetic nanodisk has been discussed in detail. Our discussion is consistent with recent numerical simulations \cite{2007Ultrafast_Toggle_Switching_of_Vortex_Cores, 2007YLiu}.

We believe that the topological current theory used in this paper may shed a light on theoretical study of the magnetic system directly from topological viewpoint.

\begin{acknowledgments}
The authors greatly acknowledge Yaowen Liu for stimulating and fruitful
discussions. This Work was supported by National Natural Science Foundation of China under Grant No. 11275119 and by Ph.D. Programs Foundation of Ministry of Education of China under Grant No. 20123108110004.
\end{acknowledgments}

\end{document}